# Explaining Network Intrusion Detection System Using Explainable AI Framework

Shraddha Mane[1] and Dattaraj Rao[2]

Persistent Systems Limited, India
`shraddha_mane@persistent.com, dattaraj_rao@persistent.com`

**Abstract.** Cybersecurity is a domain where the data distribution is constantly changing with attackers exploring newer patterns to attack cyber infrastructure. Intrusion detection system is one of the important layers in cyber safety in today's world. Machine learning based network intrusion detection systems started showing effective results in recent years. With deep learning models, detection rates of network intrusion detection system are improved. More accurate the model, more the complexity and hence less the interpretability. Deep neural networks are complex and hard to interpret which makes difficult to use them in production as reasons behind their decisions are unknown.

In this paper, we have used deep neural network for network intrusion detection and also proposed explainable AI framework to add transparency at every stage of machine learning pipeline. This is done by leveraging Explainable AI algorithms which focus on making ML models less of black boxes by providing explanations as to why a prediction is made. Explanations give us measurable factors as to what features influence the prediction of a cyber-attack and to what degree. These explanations are generated from SHAP, LIME, Contrastive Explanations Method (CEM), ProtoDash and Boolean Decision Rules via Column Generation (BRCG). We apply these approaches to NSL-KDD dataset for intrusion detection system (IDS) and demonstrate results.

## 1 Introduction

Modern intrusion detection systems leverage Machine Learning (ML) to correlate network features, identify patterns in data and highlight anomalies corresponding to attacks. Security researchers spend many hours understanding these attacks and trying to classify them into known kinds like port sweep, password guess, teardrop, etc. However, due to the constantly changing attack landscape and emergence of advanced persistent threats (APT), hackers are continuously finding new ways to attack systems. Hence a static list of classification of attacks will not be able to adapt to new and novel tactics adopted by adversaries. Also, ML systems typically end up being black boxes and do not do a very good job explaining why they flag certain network traffic. Recent research in explainable AI [1] [2] has led to explainable AI (XAI) as a dedicated field of study to explain the reasoning behind predictions made by ML models. Hence, we propose an explainable AI [3] framework along with intrusion detection system which would help analyst to make final decision.

For example, if we classify an attack as a 'guess password' and provide evidence that this is because the number of hot indicators is 1 and the source bytes is around 125 – it gives much better visibility to the security analyst on why the alert is flagged. Also, if the explanation matches the domain knowledge the analyst can easily approve it with confidence. Moreover, for a new attack when the anomaly is flagged and explanation given, the analyst may decide if this is a new pattern of unknown attack and possibly capture this as a rule [4] in a system like Zeek (formerly Bro-IDS). This analysis may eventually be combined with a natural language generating system that can provide concrete English statements explaining the alert and reasons for it.

It is seen that Deep Neural Networks (DNN) [5] outperforms ML algorithms. As per accuracy vs interpretability graph, as accuracy of the models increases, interpretability decreases. Hence less interpretability of DNN is one of the reasons why they are not deployed for real use cases such as bank loan approval. With the help of explainable AI techniques, we can convert untrustworthy model into trustworthy.

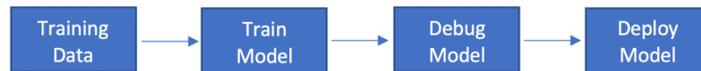

**Fig. 1.** Steps involved in ML pipeline

Above are the different steps involved in the ML pipeline. At every stage we can introduce explainable AI algorithms to get insights. Explanations generated using training data help to identify if there any biases present in the data, which can be modified before training the model. Explanations generated by using trained model can help to understand if model has learned any inappropriate rule or pattern, by retuning the model or reselecting the features better model can be obtained. Explanations which are instance specific can help to debug the model and also useful after model deployment. Thus, at every stage in ML pipeline these explanations make it easier to design and deploy best network intrusion detection system. Figure 2 shows how we can generate explanations for every type of user.

When any machine learning model is deployed into production, three types of users or consumers comes into a picture. First is data scientist who evaluates the ML model after training completes. Second is analyst - a person who takes the final decision based on model's prediction and third is end user or customer who wants to know the reason for model's prediction. Consider machine learning model for loan approval in a bank. First data scientist will design the model and before deployment he/she will check the model performance and confirm with domain expert. Loan officer will make final decision to approve or reject loan based on model's output and applicant will be keen to know reason for loan rejection or approval. For all of these users explainability is the solution.

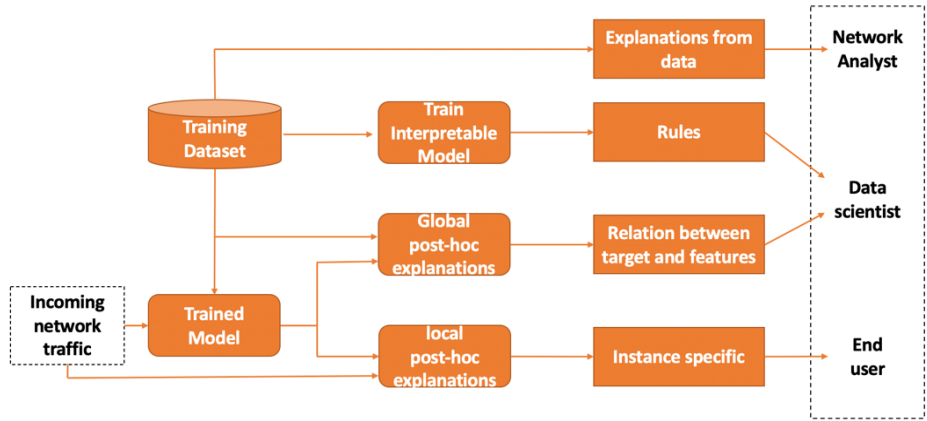

**Fig. 2.** Schematic of the explainable AI framework

For data scientist we present explainable AI methods such as SHAP [6] [7], BRCG [8] which provides global understanding of model's behaviour. For Analyst, Protodash [9] method is used, it can show samples from training dataset which are similar to given sample, what are similarities and differences between them etc. And for end user LIME [10] [11], SHAP, CEM [12] methods for local explanations are used which help to understand which features in the input instance are contributing in model's final decision and how model's decision can be changed by tweaking which feature values by what amount.

## 2    Analysis of the Dataset

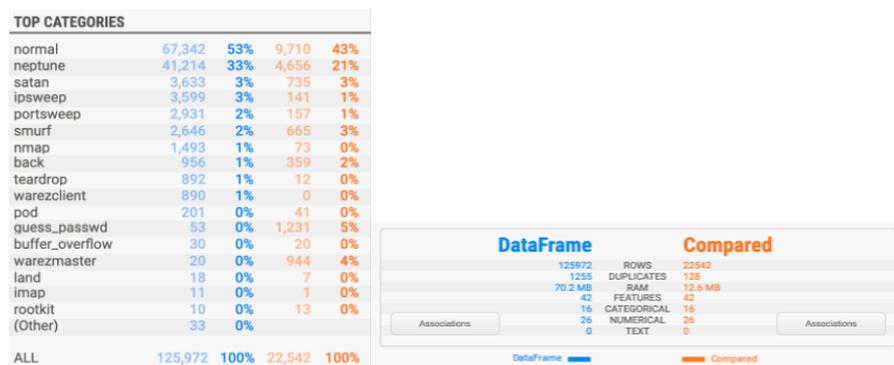

**Fig. 3.** Basic overview and attack distribution in NSL-KDD dataset

There are many datasets [13] available for IDS. We use the widely popular NSL-KDD dataset [14] [15] which contains several popular network intrusion attacks with key features captured using a network flow analysis tool and then labelled manually. Figure 3 shows comparison of Training (DataFrame) and Test (Compared) datasets volume in the NSL-KDD dataset. This analysis is done using the SweetViz [16] tool.

## 3   Classification Model

### 3.1   Preprocessing

NSL KDD data set has 42 features which consists of 3 categorical features, 6 binary features, 23 discrete features and 10 continuous features. We keep all features same except categorical features. We use one hot encoding to convert categorical features into binary features. For categorical data one hot encoding is preferred over ordinal/integer encoding as integer encoding introduces natural order relationship between categorical variables which is not true. Model may learn that ordering resulting in poor performance. Thus, number of features increase from 42 to 122. We normalized the data using min-max normalization [17] method before fitting the model on it.

### 3.2   Model Architecture

Dataset is classified into two classes – normal and attack. Any data point that is not labeled as 'normal' is treated as an 'attack'. We trained binary classification model using fully connected deep neural networks [5]. Data from 'KDDTrain+.txt' is used to train the model and for evaluation we have used 'KDDTest+.txt'.

In this experiment, fully connected network with RELU activation is used. There are 3 hidden layers in the network. Hidden layers contain 1024, 768 and 512 neurons respectively. Input dimension is 122 and output dimension is 2. Two neurons (with softmax activation) are preferred over one neuron (with sigmoid activation) in the output layer, as it provides better representation of SHAP explanations. The Structure of the classifier is – [122,1024,768,512,2].

Keras is used to build network intrusion detection model. Adam optimizer is used with learning rate of 0.01 and 100 epochs. Dropout is also used. The classifier mentioned above performs well. Following are the results.

| Accuracy | Precision | Recall | F1 |
|----------|-----------|--------|-------|
| 0.824    | 0.964     | 0.713  | 0.820 |

**Table 1.** Results on KDDTest+ dataset

## 4   Generating Explanations

Explainable AI has generated many interests in data science community from last few years. It is one of the hot research topics, many tools and libraries are releasing day by day to open up the black box models. However, there are no standard performance

5—measure metrics available to compare the performance of these methods. There is no single method of explainability which is superior than others. There are many ways to generate explanations [18] for ML models – model specific Vs model agnostic, local vs global, intrinsic vs post-hoc etc. Therefore, in this paper we have used more than one explainability techniques [19] to explain the deep learning model.

LIME is useful to generate local explanations for a particular instance which is local linear approximation for model's behavior. If we consider model's decision function, it may be very complex globally, but if we consider it around a particular instance it is easy to approximate using perturbation of samples. Linear model can be fitted around perturbed samples which will give us model insights locally.

Game theory approach is used in SHAP, it generates global and local explanations. Game theory consists of game and players, in our case game is to reproduce the outcome of the model and players are features used to train the model provided we have trained ML model with us. SHAP measures contribution made by each feature in the prediction made by the model. To determine the feature importance, outcome of every possible combination of features is considered. If there are 'n' features, then SHAP trains '2n' distinct predictive models. Dataset used is same in all models only change is number of features considered. The gap between predictions made by these models will help to calculate overall importance of the feature.

IBM research has developed open source toolkit named AI explainability 360 [20](AIX 360) for interpretability and explainability of machine learning models as well as data. It supports eight different algorithms for explainability. We have used three algorithms to explain deep neural network, according to their use in various steps of AI modelling pipeline.

ProtoDash algorithm uses training data and provides exemplar based explanations for summarizing dataset, it also explains prediction made by the model. BRCG trains interpretable model by using supervised learning algorithm for binary classification. It learns Boolean rules from data which are combination of simple OR of ANDs rules or AND of ORs rules. CEM is used to generate local explanations on trained model for a particular instance from the dataset. It finds what is minimally sufficient (PP – Pertinent Positive) and what should be necessarily absent (PN – Pertinent Negative) to maintain the original classification.

| Algorithm | ProtoDash | Boolean Decision Rules via Column Generation (BRCG) | Contrastive Explanation Method (CEM) |
|---|---|---|---|
| Use | Explanations from training data | Train interpretable Model | Local post-hoc explanations |

**Table 2.** AIX 360 algorithms

## 5      Results

SHAP summary plot is global explanation of a model which combines feature importance with feature effect. Shapely value for a feature and particular sample is represented by a point on a summary plot. Features are on Y axis and shapely values are on X-axis. Colors are used to represent low/high values. Features are arranged according to their importance, top feature in the summary plot is most important whereas bottom one is the least.

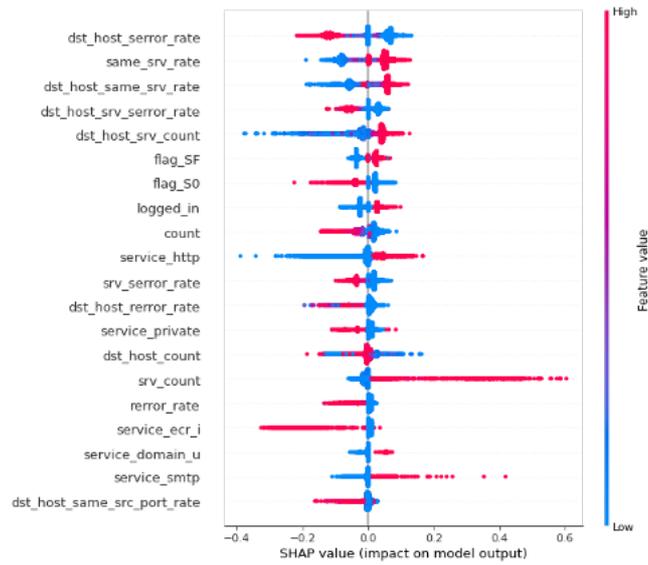

**Fig. 4.** SHAP summary plot – high value of 'same_srv_rate' increases the probability of attack whereas high value of 'dst_host_serror_rate' decreases the probability of attack

SHAP global explanations are drawn by considering complete/partial dataset. SHAP local explanations consider only specific instance at a time and generates explanation, it shows which feature values are taking decision towards positive and which are taking towards negative. Figure 5 shows local explanation where probability of the output being attack is 1.00 and features along with their values are shown below such as 'dst_host_same_srv_rate', 'same_srv_rate', 'service_private' and so on. Features pushing the prediction higher are shown in red and those pushing predictions to lower are shown in blue. Note that these explanation changes as we change the input instance.

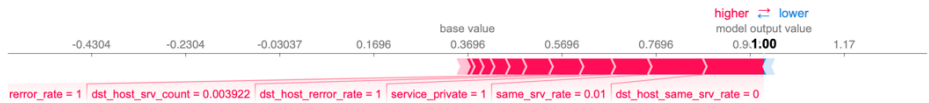

**Fig. 5.** SHAP force plot used for local explanations to explain a particular instance where output probability of 'attack' is 1 and shows features contributing in decision

Figure 6 below shows the SHAP force plot for group of points from test dataset. We combined 50 points from each category - normal and 3 types of attacks and plotted a force plot for this shown below. This is basically created by taking multiple force plots for a single instance (shown in figure 5) rotating them by 90 degrees and stacking them horizontally. We see a clear separation on type of attack defined by the explanations.

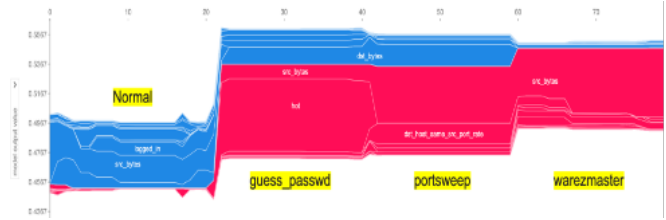

**Fig. 6.** SHAP force plot for 4 types of data points in NSL-KDD dataset

This shows how the influencing feature vary over the group of points.

We have trained DNN [5] to identify if network traffic is 'normal' or 'attack'. We can summaries the model by using rules. We have used BRCG to extract rules out of the data. Rules are as follows:

**Predict Y=1(Attack) if ANY of the following rules are satisfied, otherwise Y=0 (Normal):**
  · **wrong_fragment > 0.00**
  · **src_bytes <= 0.00 AND dst_host_diff_srv_rate > 0.01**
  · **dst_host_count <= 0.04 AND protocol_type_icmp**
  · **num_compromised > 0.00 AND dst_host_same_srv_rate > 0.98**
  · **srv_count > 0.00 AND protocol_type_icmp AND service_urp_i not**

It is interesting to note the performance of the BRCG algorithm, it is as follows:
  **Training Accuracy = 0.9823**
  **Test Accuracy = 0.7950**

Which means by applying just these rules one can get ~80% accurate results on test data.

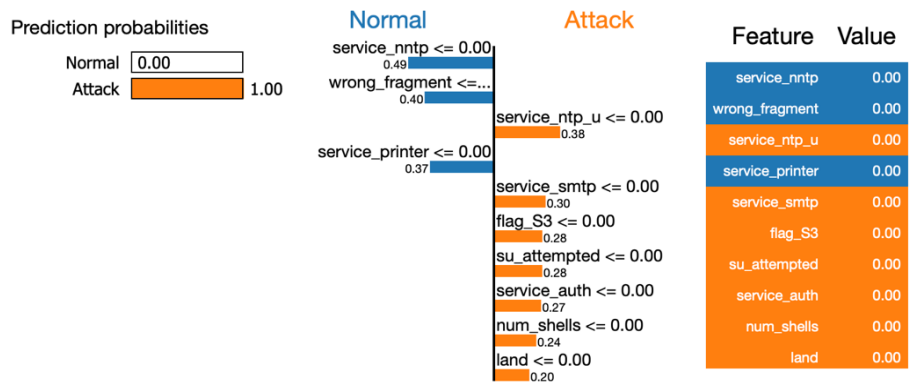

**Fig. 7.** Explaining individual prediction of deep learning classifier using LIME

LIME generates local explanations. In figure 7 explanation is shown to determine if classification result is 'Normal' or 'Attack' along with probability and original instance values. Colors are used to highlight which features contributes to which class. Features in orange colour contributes to 'attack' and blue contributes to 'normal' category.

|  | 0 | 1 | 2 | 3 | 4 |
|---|---|---|---|---|---|
| duration | 0 | 0 | 0 | 0 | 0 |
| src_bytes | 0 | 7.47846e-07 | 0 | 0 | 0 |
| dst_bytes | 0 | 0 | 0 | 0 | 0 |
| land | 0 | 0 | 0 | 0 | 0 |
| wrong_fragment | 0 | 0 | 0 | 0 | 0 |
| ... | ... | ... | ... | ... | ... |
| flag_S3 | 0 | 0 | 0 | 0 | 0 |
| flag_SF | 0 | 1 | 0 | 0 | 0 |
| flag_SH | 0 | 0 | 0 | 0 | 0 |
| Class | Attack | Attack | Attack | Attack | Attack |
| Weight | 0.935025 | 3.00021e-05 | 0.000150011 | 0.0255018 | 0.0392928 |

|  | 0 | 1 | 2 | 3 | 4 |
|---|---|---|---|---|---|
| duration | 1.0 | 1.00 | 1.0 | 1.0 | 1.0 |
| src_bytes | 1.0 | 0.08 | 1.0 | 1.0 | 1.0 |
| dst_bytes | 1.0 | 1.00 | 1.0 | 1.0 | 1.0 |
| land | 1.0 | 1.00 | 1.0 | 1.0 | 1.0 |
| wrong_fragment | 1.0 | 1.00 | 1.0 | 1.0 | 1.0 |
| ... | ... | ... | ... | ... | ... |
| flag_S1 | 1.0 | 1.00 | 1.0 | 1.0 | 1.0 |
| flag_S2 | 1.0 | 1.00 | 1.0 | 1.0 | 1.0 |
| flag_S3 | 1.0 | 1.00 | 1.0 | 1.0 | 1.0 |
| flag_SF | 1.0 | 0.08 | 1.0 | 1.0 | 1.0 |
| flag_SH | 1.0 | 1.00 | 1.0 | 1.0 | 1.0 |

**Table 3a.** Similar instances predicted as attack **3b.** Use of weights to show similarity

The person who takes final decision based on model's output can get understanding of model's decision if we show instances from the training dataset which are similar in different ways to test instance we want to understand. We considered first instance from test dataset for which model prediction is 'attack'. Table 3a shows similar instances from training data, similarity is indicated by the weight mentioned in last row.

It also provides human friendly explanations showing feature values in terms of weight. More the weight, more the similarity. Above two tables- table 3a and table 3b represent five closest instances to the test instance. Based on the weights mentioned, we can see that instance under column 0 is the most representative of test instance as weight it 0.93. These tables would help the analyst to take final decision more confidently.

For end users, ML models should be transparent. They should get answers to their all queries such as why model made certain decision, which factors led to this decision, by making what changes model's decision can be changed etc. CEM algorithm helps us to answer all these end user questions.

We considered one particular instance where prediction made was 'normal', CEM shows us how decision can be changed by making minimal changes in the feature values.

CEM can also highlight minimal set of features along with their values that would maintain the prediction made by the model. Looking at the statistics of explanations given by the CEM algorithm over bunch of applicants, one can get insight into what minimal set of features play important role. It is also possible to get values of these features for every type of attack.

```
Sample: 2
prediction(X) [[1.00e+00 3.15e-10]] Normal
prediction(Xpn) [[0.18 0.82]] Attack
```

|  | X | X_PN | (X_PN - X) |
|---|---|---|---|
| duration | 3.4653e-05 | 0.02 | 0.02 |
| hot | 0 | 0.12 | 0.12 |
| dst_host_serror_rate | 0 | 0.03 | 0.03 |
| Class | Normal | Attack | NIL |

```
PP for Sample: 5
Prediction(Xpp) : Normal
Prediction probabilities for Xpp: [[0.55 0.45]]
```

|  | X | X_PP |
|---|---|---|
| num_root | 0 | 0.02 |
| count | 0.00782779 | 0.03 |
| srv_count | 0.00782779 | 0.06 |
| diff_srv_rate | 0 | 0.02 |
| dst_host_count | 0.607843 | 0.01 |
| Class | Normal | Normal |

**Table 4.** Pertinent negative and pertinent positives for an instance

First table shows that if we change feature values of only three features – duration, hot, dst_host_serror_rate from 0, 0, 0 to 0.02, 0.12, 0.03 respectively by keeping rest all feature values same, the classifier decision changes from 'normal' to 'attack'. Second table shows minimum values required for the features to maintain same decision of the model. It shows five parameters num_root, count, srv_count, diff_srv_rate and dst_host_count along with their original value and minimum value required to model to make same decision.

## 6    Conclusion

Nowadays, Intrusion detection systems which uses machine learning algorithms are untrusted as reasons behind their decisions are unknown. We presented a framework where local and global explanations help to get important features in decision making, instance-wise explanations and relation between features and outcome of the system. By looking at the explanations Data scientists can understand what patterns model has learned, if learned patterns are wrong then data scientist can choose different set of features or make changes in the dataset so that model learns appropriately.

Network analyst makes the final decision by considering model's prediction. Along with prediction, we provide explanations where similar instances are shown for which prediction of the model is same as the test instance so that network analyst understands similarities between them and can make final decision. Explanations provided for end users help them to understand which features are contributing in decision making with what weightage. So that they can alter values of features to change the model's decision.

Thus, presented framework provides explanations at every stage of machine learning pipeline which are suited for different users in network intrusion detection system.